\documentclass[twocolumn,showpacs,preprintnumbers,amsmath,amssymb]{revtex4}

\usepackage{graphicx}
\usepackage{dcolumn}
\usepackage{bm}
\usepackage{times}

\begin{document}
\baselineskip 10.8pt \preprint{APS/123-QED}

\title{Decoherence of the flux-based superconducting qubit in an integrated circuit environment}

\author{Jonathan L. Habif and Mark F. Bocko}
\affiliation{SDE Laboratory, Department of Electrical and Computer
Engineering, University of Rochester, Rochester, NY 14627}
\date{\today}

\begin{abstract}
Superconducting, flux-based qubits are promising candidates for
the construction of a large scale quantum computer.  We present an
explicit quantum mechanical calculation of the coherent behavior
of a flux based quantum bit in a noisy experimental environment
such as an integrated circuit containing bias and control
electronics.  We show that non-thermal noise sources, such as bias
current fluctuations and magnetic coupling to nearby active
control circuits, will cause decoherence of a flux-based qubit on
a timescale comparable to recent experimental coherence time
measurements.
\end{abstract}

\maketitle
\section{ INTRODUCTION}
The possibility of observing quantum coherent behavior in
macroscopic devices, such as superconducting circuits, was first
suggested by A. J. Leggett in the mid 1980's \cite{1}.  Since
then, experimentalists have attempted to observe such behavior
\cite{2,3} and, more recently, to extend the idea to a full scale
quantum computer capable of executing quantum algorithms such as
Shor's algorithm \cite{4}.  Superconducting qubit types include
charge qubits: created by a superposition of the number of
electrons on a superconducting island \cite{5}; phase qubits: the
superposition of energy states in a single Josephson junction
\cite{6}; and flux qubits: the superposition of the quantity of
flux threading a superconducting ring interrupted by a Josephson
junction \cite{7}. Quantum mechanical behavior has been verified
in all three of these qubit systems, and experiments have shown
that each system can be prepared in quantum superpositions. The
coherence times which result from these environmental
interactions, however, are far shorter than those predicted by
current theories of open quantum systems.

Non-thermal sources of noise, such as fluctuations from room
temperature laboratory control and measurement equipment and
active circuit elements operating in the vicinity of the quantum
device will contribute to the decoherence of the qubit in ways
that are unique to different experimental configurations and
methods.  In this work we calculate the coherence time of a
flux-based quantum bit exposed to classical noise sources
consistent with those that would be found in an integrated circuit
environment.  We explicitly calculate the evolution of the qubit
wavefunction under the influence of a Hamiltonian including
environmental noise and infer decoherence times via an ensemble
average of such calculations.

The future of solid state quantum computation will likely involve
the integration of quantum bits onto a monolithic circuit also
containing classical electronics used for quantum state
preparation, manipulation and readout \cite{4,8}.  For a
large-scale quantum computer the classical control electronics
will most likely be digital and be implemented in a well
understood, robust technology such as superconducting rapid single
flux quantum (RSFQ) technology.  RSFQ logic is an integrated
circuit family
\begin{figure}[tb]
\centering
\includegraphics[width=8.6cm]{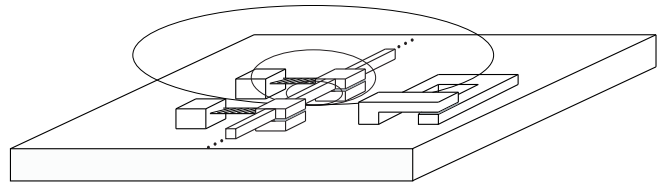}
\caption{A RSFQ circuit (left) physically near and inductively
coupled to a rf-SQUID qubit (right).  The RSFQ circuit passes
digital bits as quantized units of magnetic flux.  The physical
proximity of the active circuit will destroy the coherent behavior
of the rf-SQUID qubit.  The elliptical lines surrounding the RSFQ
inductor are the field lines created by a passing data pulse.}
\label{fig:integ_cct}
\end{figure}
that uses single flux quantum (SFQ) magnetic pulses as data bits.
A diagram of a likely physical layout of an integrated circuit is
illustrated in Fig. \ref{fig:integ_cct}, where RSFQ digital
circuitry is placed near by and inductively coupled to a rf-SQUID
qubit. Unlike charge and phase based Josephson qubits, the flux
based rf-SQUID qubit is inductively coupled to the environment
making it susceptible to the effects of stray magnetic fields. The
quantum dynamics of the rf-SQUID is exceedingly sensitive to the
applied external magnetic field.  The mutual inductance of the
rf-SQUID qubit facilitates simple coupling procedures between
qubits and classical circuitry, but also leaves the qubit
vulnerable to unwanted coupling to active circuit elements
integrated on the same chip.  Although one clearly will attempt to
shield the qubit from such stray fields these measures cannot be
perfect. Furthermore, in order to be of any practical use, the
flux qubits can not be completely isolated. Studies have been
performed evaluating the effect of mutual inductive coupling in
standard CMOS circuit technology \cite{9}.  Due to the robustness
of digital logic, classical circuitry has a high level of immunity
to these types of effects, contrary to quantum coherent
technologies, which will be extremely sensitive to noise coupled
through a mutual inductance.

Though quantum dynamics of the rf-SQUID have been verified, the
macroscopic quantum coherent (MQC) oscillations of flux, predicted
by A.J. Leggett \cite{1} have not yet been directly observed due
to the lack of sufficiently sensitive and fast means of
measurement. By integrating the entire MQC experiment onto a
monolithic circuit we believe it will be possible to create the
necessary electronics to conduct the experiment.  Furthermore,
such chips used to perform the necessary experimental procedures
can be used to explore decoherence mechanisms arising from qubit
coupling to classical noise sources.

The calculation performed in the following sections models the
decoherence of a rf-SQUID qubit integrated with active, classical
circuitry.  The effect of the classical circuits on the qubit is
modeled by the inclusion of fluctuating terms in the Hamiltonian
of the rf-SQUID.  The classical sources of noise are characterized
by Gaussian random variables with a finite bandwidth and
predetermined power spectral density (PSD).  The predicted qubit
coherence times depend upon characteristics of the classical noise
that may be measured experimentally.

\section{The Decoherence Calculation}
Decoherence can most easily be defined as the deviation of the
behavior of a quantum mechanical system from that predicted by the
Schrödinger equation for the closed quantum system. Traditionally,
decoherence is defined in the context of the density matrix
($\rho$) representation where loss of coherence in a quantum
system is indicated by the suppression of the off-diagonal
elements of the density matrix.  In a given basis,
\begin{equation}
\rho(t) = \frac{1}{N}\sum_{i=1}^{N}|\psi_{i}(t)\rangle \langle
\psi_{i}(t)|, \label{eq:densitymatrix}
\end{equation}
where $N$ is the number of individual systems composing the
ensemble about which quantum statistics are being described and
$|\psi_{i}(t)\rangle$ is the state of the $i^{th}$ system.  When
the off-diagonal elements of $\rho$ are completely suppressed
evaluation of the Schr\"{ö}dinger equation for the closed system
ceases to be an accurate predition of the evolution of the state
$|\psi \rangle$ of the open system.  Decoherence models have been
created to predict the time scales on which non-deterministic
degrees of freedom of the environment become entangled with the
qubit thereby destroying the coherence of the quantum system
\cite{10}. There is a large body of work on the decoherence of a
spin-1/2 system coupled to a reservoir of harmonic oscillators
\cite{11,12}. To our knowledge, however, an explicit,
time-dependent calculation of the evolution of a flux-based qubit
including the effects of a noisy environment has not been
completed.

Two distinguishable decoherence mechanisms contribute to the
suppression of the off-diagonal elements of $\rho$.  Relaxation is
associated with an increase or reduction of the expectation value
of the energy.  Dephasing is an adiabatic process whereby the
phase of the system wavefunction becomes randomized.  Dephasing
typically occurs on a much shorter timescale than relaxation
\cite{13}.

The Hamiltonian for the rf-SQUID, derived in \cite{1}, is given
by,
\begin{equation}
H = \frac{-\hbar^{2}}{2C} \frac{\partial^{2}}{\partial \Phi^{2}} +
\frac{\left( \Phi - \Phi_{x} \right)^{2}}{2L} - \frac{I_{c}
\Phi_{0}}{2 \pi} \cos \left( \frac{2 \pi \Phi}{\Phi_{0}} \right).
\label{eq:Hamiltonian}
\end{equation}
In this equation $C$ is the capacitance of the Josephson junction,
$\Phi_{x}$ is the magnetic flux applied to the device externally
and $L$ is the loop inductance. The independent variable, $\Phi$
is the flux threading the superconducting loop.  The first term
represents the kinetic energy of the SQUID, while the second and
third terms constitute the potential energy of the superconducting
inductor and Josephson junction, respectively.

The qubit potential has two independent degrees of freedom, the
height of the barrier separating the two minima, and the relative
depth of the two minima.  The Hamiltonian can be rewritten in a
form similar to that of a two-state system as,
\begin{equation}
H_{tss} = -\frac{\hbar \Delta}{2} \sigma_{x} + \frac{\epsilon}{2}
\sigma_{z}, \label{eq:twostate}
\end{equation}
where $\Delta$ is the tunnelling matrix element, and $\epsilon$ is
the difference in energy between the ground states of the two
wells. The operators $\sigma_{x}$ and $\sigma_{z}$ are Pauli
matrices.  In principle these two degrees of freedom are
separable, and can be considered independently.  In a laboratory
setting, however, they are subject to the same environmental
influences and should be considered simultaneously when examining
decoherence.  The flux states of the system correspond to the
eigenstates of $\sigma_{z}$.  In this basis fluctuations in the
barrier height are referred to as $\sigma_{x}$ fluctuations
because they modulate the energy level spacing and fluctuations in
the relative depths of the wells are called $\sigma_{z}$
fluctuations.  When the flux bias deviates from $\Phi_{0}/2$ by
greater than $10^{-4} \Phi_{0}$, the energy and flux basis states
are nearly the same. However, when the system is flux biased at
exactly $\Phi_{0}/2$ the energy bases are non-local in flux; there
is finite probability of finding the flux in either well.
Traditionally, $\sigma_{z}$ fluctuations are considered to be the
most destructive to the coherence of a system.  In a typical
experiment[2], the single rf-SQUID junction is replaced by a
double junction loop of small self-inductance. This added
inductance will promote coupling between the environmental
fluctuations and $\sigma_{x}$ degree of freedom in the system. The
contribution to dephasing from $\sigma_{x}$ and $\sigma_{z}$
degrees of freedom will depend on the ratio of the
self-inductances of the junction loop and SQUID loop,
respectively.  Reducing the junction loop inductance will allow
$\sigma_{x}$ dephasing to be ignored.

As stated earlier, the calculation performed in this paper is an
explicit solution of the time-dependant Schrödinger equation.
Previous calculations of the decoherence of a rf-SQUID have
focused on modeling the interaction of the environment as a
continuous weak measurement \cite{14} or by reducing the problem
to a two-state system and using the spin-boson formalism
\cite{12,15}. The shortcomings of these models are that they
approximate the effect of a general environment on the system, but
do not accurately reflect the dominant sources of noise in a
circuit environment. For example, the spin-boson formalism reduces
the qubit to a system with only two energy levels and models the
interaction between the qubit and environment as bilateral, in
that the state of the qubit affects the state of the environment.
In an integrated circuit environment the back action of the qubit
on the sources of the fluctuating fields is insignificant.

The illustration in Fig. \ref{fig:integ_cct} depicts an RSFQ
integrated circuit, consisting of resistors and superconducting
inductors and Josephson junctions near a rf-SQUID qubit.  The
magnetic data pulses that propagate through the RSFQ circuit will
inductively couple flux to the qubit perturbing the otherwise
constant external flux bias; it is necessary to determine the
magnitude of these decoherence causing fluctuations.  Stray
magnetic flux coupled to the qubit from the RSFQ circuitry is
determined by the magnitude of the current pulses in nearby RSFQ
circuits and the mutual inductance ($M$) between the circuits and
qubit.  A typical estimate for $M$ when the circuit and qubit are
$20 \mu m$ apart is on the order of $10 fH$.  A SFQ data pulse
passing through a superconducting inductor of typical size in RSFQ
circuits has a peak amplitude of approximately $150 \mu A$ and a
duration of $2-4 ps$. With these parameters the excess flux
coupled to the qubit is $720 \mu \Phi_{0}$.  An analysis of the
decoherence effects in such a situation follows.

The effect of a large number of nearby RSFQ circuits, each passing
SFQ pulses at seemingly random times from the point of view of the
qubit, may be represented by introducing a random component to the
flux in the vicinity of the qubit.  Thus, our calculation was
performed by introducing noise terms directly into the
Hamiltonian, Eq. \ref{eq:Hamiltonian}, and then solving the
time-dependant Schrödinger equation with the random Hamiltonian.
The rf-SQUID can be perturbed by the flux environment in two ways:
the external flux bias applied to the qubit can deviate from
$\Phi_{0}/2$, or magnetic flux can couple to the Josephson
junction thereby changing the effective $I_{c}$.  As shown earlier
the influence of the fluctuations of the applied flux bias,
$\sigma_{z}$ noise, dominates the effect of qubit junction
critical current fluctuations, the $\sigma_{x}$ noise. Therefore a
random term was added to in the Hamiltonian and the evolution of
the qubit was computed. The initial wavefunction, $|\Psi_{0}
\rangle$, used in the calculation was determined by the ground
state of the system with slightly less than $\Phi_{0}/2$ applied
flux. At milliKelvin temperatures the system will completely relax
to this state. $|\Psi_{0} \rangle$ is then projected onto the
basis functions of the symmetric double well potential. The state
of the system is nearly a pure superposition of the ground and
first excited state of the symmetric potential. There are,
however, excitations of higher lying states that are small but
nonetheless included in the calculation.

The time evolution of the system is determined using a time
varying propagator operating on the wavefunction,
\begin{equation}
|\psi_{m}(t+\Delta t) \rangle = exp(iH_{m}(t+\Delta t)\Delta
t)|\psi_{m}(t) \rangle, \label{eq:propagator}
\end{equation}
where $H_{m}(t+\Delta t)$ is the Hamiltonian of the system with
$\Phi_{x}(t+\Delta t)$ and $I_{c}(t+ \Delta t)$ ; we also chose
$\Delta t$ to be $100 ps$, much shorter than the period of the MQC
oscillations. At each time step the wave function $|\psi_{m}(t)
\rangle$ is projected onto the basis states of the new Hamiltonian
$H_{m}(t+\Delta t)$, which are determined before the calculation
and stored in a look up table, and then evolved through time
$\Delta t$ to produce the new wavefunction, $|\psi_{m}(t+\Delta t)
\rangle$. The result for short time-scales is a randomization of
the phase of the coherent oscillations of the flux.  On a longer
time-scales the populations of the energy levels will deviate from
those of $|\Psi_{0} \rangle$. Since the loss of phase coherence
occurs on a shorter time-scale that is the dominant form of
decoherence and thus will be the subject of this work.

In order to calculate the dephasing time of the MQC oscillations
in the fluctuating potential the simulation described above is
repeated many times, and the separate results are averaged as
shown in Eq. \ref{eq:densitymatrix}.  An experimental technique
similar to the one described here was used in \cite{16} to measure
the phase stability of classical oscillators.  For our calculation
we used $N=50$. For each trial the phase of the oscillations
becomes unpredictable after some period of time.  Thus averaging
over a large number of random phases leads to the vanishing of the
off-diagonal elements of the density matrix in Eq.
\ref{eq:densitymatrix}.

A set of $N$ evolutions was performed and summed for each value of
\begin{figure}[tb]
\centering
\includegraphics[width=8.6cm]{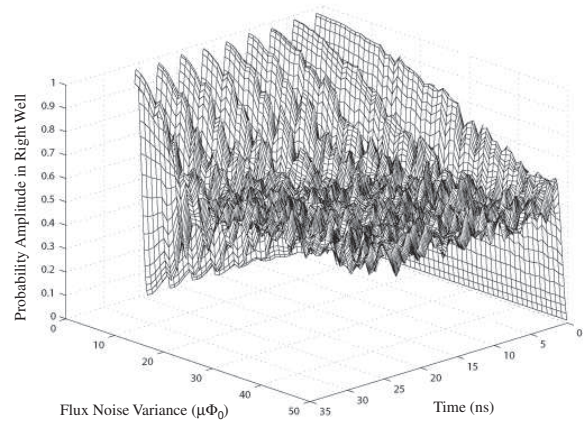}
\caption{A surface plot of the data obtained from the calculation
of decoherence of the rf-SQUID qubit.  As the variance of the flux
noise influencing the qubit is increased, the coherent
oscillations of flux last for shorter amounts of time.}
\label{fig:surfplot}
\end{figure}
flux noise variance, and the results are plotted in the surface
plot of Fig. \ref{fig:surfplot}.  The bandwidth of the noise in
Fig. \ref{fig:surfplot} is $4 GHz$. To quantify the dephasing time
resulting from each value of flux noise variance, the damped
sinusoid resulting from the set of N evolutions was fit to a
function of the form,
\begin{equation}
P(t) = exp(-D_{\phi}t)cos(\omega_{0}t + \pi), \label{eq:dampfit}
\end{equation}
where $\omega_{0}$ is the resonant frequency of the tunnelling
flux, which is also equal to the energy splitting between the
ground and first excited states.  The value of the dephasing time
constant, $D_{\phi}$, for each value of flux noise variance was
\begin{figure}[b!]
\centering
\includegraphics[width=8.6cm]{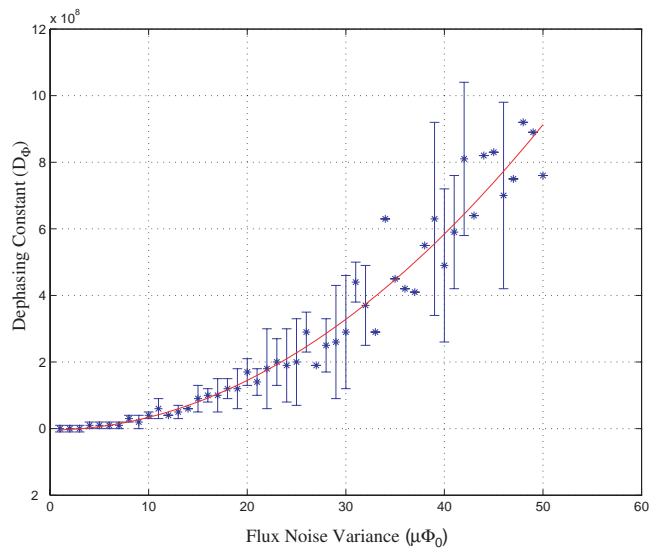}
\caption{A plot of the dephasing rate as a function of flux noise
variance.  The bandwidth of the noise signal used for this data is
$4 GHz$.  Error bars are provided for selected data points
indicating the confidence of the fit.} \label{fig:dataplot}
\end{figure}
determined using a least squares fit to the data obtained in Fig.
\ref{fig:surfplot}.  Fig. \ref{fig:dataplot} shows the plot of
$D_{\phi}$ as a function of the flux noise variance,
$\sigma_{\phi}$, applied to the qubit. From the graph in Fig.
\ref{fig:dataplot} the relation between $D_{\phi}$ and the
variance is $D_{\phi} = 3.63 \cdot 10^{-4} \left( \sigma_{\phi}
\right)^{2}$ (1/ns), where $\sigma_{\phi}$ is in units of $\mu
\Phi_{0}$. Since $\sigma_{\phi}^{2} = \int_{-\infty}^{\infty}
S_{\Phi}(\omega) d \omega$, $D_{\phi}$ is linearly proportional to
the amplitude of the power spectral density of the environmental
flux noise at a given bandwidth. Fig. \ref{fig:surfplot} shows
that if the flux noise variance is increased above approximately
$10 \mu \Phi_{0}$ , then the coherence of the qubit lasts less
than $20 ns$.  From the physical scenario described above, it is
evident that $M$ must be reduced by at least two orders of
magnitude, to below $.1 fH$, in order for the excess flux to be
reduced to these levels and for acceptable coherence times to be
achieved.

An analysis was also performed to measure the dependence of
$D_{\phi}$ on the bandwidth of the noise signal.  It was seen that
if the noise is bandlimited to a cutoff frequency, $\omega_{c}$,
below $\omega_{0}$, then $D_{\phi}$ varies approximately linearly
with the noise bandwidth.  If $\omega_{c} \geq \omega_{0}$,
$D_{\phi}$ is nearly independent of the bandwidth of the noise
signal. When the noise is bandlimited so that $\omega_{c} \leq
\omega_{0}$ the frequency of the oscillations is no longer
$\omega_{0}$ due to the fact that on the timescales in question
noise bandlimited below $\omega_{0}$ can no longer be considered
white, and $\overline{\Phi_{x}} \neq \frac{\Phi_{0}}{2}$.  This
yields a noticeable increase in the coherent tunnelling frequency
of the system.  At noise frequencies greater than $\omega_{0}$ the
noise can be considered white and $\overline{\Phi_{x}} =
\frac{\Phi_{0}}{2}$ over the 35 ns time scale of the simulation.
At large noise frequencies the dephasing rate, $D_{\phi}$, is
dependent primarily on the variance of the flux noise and nearly
independent of the bandwidth of the noise. In an integrated
circuit environment, especially one operating at GHz speeds, the
bandwidth of the noise created by the circuitry will certainly
exceed that of $\omega_{0}$, which in this case is approximately
280 MHz.

\section{Conclusions}
The ultimate goal of integrating classical electronics with
coherent quantum circuitry is to efficiently prepare and
manipulate the coherent states of the qubits.  This will require
that classical circuitry, such as RSFQ, be placed in close
proximity to the quantum coherent bits in order to provide the
necessary interaction, which in this case is inductive.  Unwanted
interactions between unintentionally coupled classical circuitry
and qubits will no doubt exist and it must be determined what
effect that will have on the coherence time of the qubits.  We
have calculated that for current RSFQ technology there is a
threshold value for the mutual inductance between RSFQ circuitry
and a qubit of approximately $0.1 fH$ below which qubit coherence
times may persist for longer than $20 ns$.

We have used an explicit formulation of the Hamiltonian of the
rf-SQUID in a noisy environment in order to determine a realistic
value for the dephasing time in an integrated circuit environment.
Moreover, the parameter that accounts for the loss of coherence in
this model, the flux noise variance, is easily measured by
incorporating a dc-SQUID magnetometer in the vicinity of the
qubit.  The approach presented here for estimating the coherence
time of a rf-SQUID qubit is not intended as a general model of
decoherence for two-state systems.  Rather it provides an explicit
model from which coherence times of rf-SQUID qubits may be
estimated based upon experimentally measurable quantities.  Models
such as this one coupled with measurements of the flux noise at
the location of the qubit will help quantum computer architects
design large scale computers capable of executing the algorithms
for which they were originally intended.

\section{AKNOWLEDGEMENTS}
Supported in part by AFOSR grant F49620-01-1-0457 funded under the
Department of Defense University Research Initiative on
Nanotechnology (DURINT) program and by the ARDA.  J.L.H. is
supported by a NASA GSRP fellowship.

\end{document}